\documentclass[pra,twocolumn,showpacs,amsfonts,amsmath,floatfix]{revtex4}

\usepackage{graphicx}
\usepackage{epsfig}
\usepackage{epstopdf}
\usepackage[usenames,dvipsnames]{xcolor}

\newcommand{\lt}{\left(}
\newcommand{\rt}{\right)}
\newcommand{\lqu}{\left[}
\newcommand{\rqu}{\right]}
\newcommand{\lgr}{\left\{}
\newcommand{\rgr}{\right\}}

\newcommand{\be}{\begin{equation}}
\newcommand{\ee}{\end{equation}}
\newcommand{\ba}{\begin{eqnarray}}
\newcommand{\ea}{\end{eqnarray}}
\newcommand{\fr}{\frac}
\newcommand{\nn}{\nonumber}

\newcommand{\sse}{\subsection}
\newcommand{\ssse}{\subsubsection}

\begin{document}

\title{A direct approach to Gaussian measurement based quantum computation}

\author{G. Ferrini$^{1, 2, 3}$, J. Roslund$^1$, F. Arzani$^1$,  C. Fabre$^1$ and N. Treps$^1$}
\address{$^1$ Laboratoire Kastler Brossel, UPMC-Sorbonne Universit\'es, CNRS, ENS-PSL Research University, College de France; CC74, 4 Place Jussieu, 75252 Paris, France}
\address{$^2$ Laboratoire Mat\'eriaux et Ph\'enom\`enes Quantiques, Sorbonne Paris Cit\'e, Univ. Paris Diderot, CNRS UMR 7162, 75013, Paris, France}
\address{$^3$ Institute of Physics, Johannes-G{\"u}tenberg Universit{\"a}t Mainz, Staudingerweg 7, 55128 Mainz, Germany}

\email{giulia.ferrini@gmail.com} 
\date{\today}

\begin{abstract}

In this work we introduce a novel scheme for measurement based quantum computation in continuous variables.
Our approach does not necessarily rely on the use of ancillary cluster states to achieve its aim, but rather on the detection of a resource state in a suitable mode basis followed by digital post-processing, and involves an optimization of the adjustable experimental parameters. After introducing the general method, we present some examples of application to simple specific computations.

\end{abstract}
\maketitle

\section{Introduction}

Continuous Variable (CV) quantum computing (QC) in the measurement based setting~\cite{Nielsen2006,Gu2009} is emerging as a promising paradigm for quantum computation~\cite{yokoyama2013optical, Chen2014}.
In the standard approach, measurement based quantum computations (MBQCs) are carried out by fabricating a highly entangled resource state possessing specific quantum correlations, the cluster state, to which the state to be processed is entangled. The manipulation of the input state is then carried out by performing judiciously chosen local projective measurements on the nodes of the cluster state, thereby projecting the remaining nodes onto the desired computation result~\cite{Briegel2001,Raussendorf2001one}.

The effort in this procedure is hence devoted towards the construction of a large cluster state. The usual method to create a CV-cluster state consists in disposing of a set of squeezed states, often individually created in separate cavities~\cite{Furusawa2011, Su12, Armstrong2012}, and in transforming them in a set of entangled modes by a suitable network of beam-splitters and dephasers~\cite{vanLoock2007, Furusawa2011,Su07, Su12}. In this approach, the configuration of the network depends on the specific cluster state to be generated, and its complexity grows rapidly with the number of modes, rendering this method difficult to scale~\cite{vanLoock2007, vanLoock2008, Furusawa2011, Su07, Su12}.
In recent experiments large cluster states have been constructed with time~\cite{yokoyama2013optical} or frequency~\cite{Chen2014} encoding. The ability to perform a quantum computation (QC) on a resource state that is consumed in time opens the possibility of scaling the computation to large mode numbers. 

In this work we explore a different avenue, and we propose a new approach to MBQC that is distinct from the traditional one premised upon the explicit use of cluster states. This scheme is still based on the use of ancillary squeezed states, but is software-based and utilizes post-processing following a measurement 
as a means to discover the most suitable basis in which to express the QC result. Thus, the method directly targets a desired result. 

The result of a quantum computation is the set of outcomes of quadrature measurements on the modes of the output state after a desired unitary evolution has been performed. In this work we consider Gaussian operations, i.e the output state is
\be
\label{eq:Qcomp}
| \psi \rangle_{\text{out}} = e^{i H_G(\hat{q_i},\hat{p_i})} | \psi \rangle_{\text{in}},
\ee
where $H_G(\hat{x}_i,\hat{p}_i)$ is the Hamiltonian defining the evolution and is at most quadratic in the quadrature operators $\hat{x}_i,\hat{p}_i$ of each mode $i$. Our method works for arbitrary multimode input states $| \psi \rangle_{\text{in}}$ even when these are non-Gaussian. 
Therefore sampling from the probability distribution of  the output quadrature measurement outcomes is not necessarily 
a problem efficiently simulatable by a classical computer~\cite{Mari2012, Rahimi-Keshari2015}. 

After presenting a general formulation of CV measurement based quantum computation in Sec.\ref{se:MBQC},  in Sec.\ref{se:charac} we present the characterization of the operations that can be induced by simple measurement of the input state on a suitable basis and post-processing, without the use of ancillary squeezed states that are quantum correlated to the input. We show that there exist Gaussian operations that cannot be achieved by this trivial measurement method. The latter operations require ancillary resources to be performed.  In Sec.\ref{se:DMBQC} we present our direct MBQC method, and we show that it allows achieving examples of these non-trivial operations. We conclude in Sec.\ref{se:concl}.

\section{Gaussian Measurement Based Quantum Computing}
\label{se:MBQC} 

We start by reformulating Gaussian MBQC in a general framework. This formulation encompasses both the standard approach based on the use of cluster states, as well as our direct approach. 

\begin{figure}
\centering
\includegraphics*[width=1.15\columnwidth]{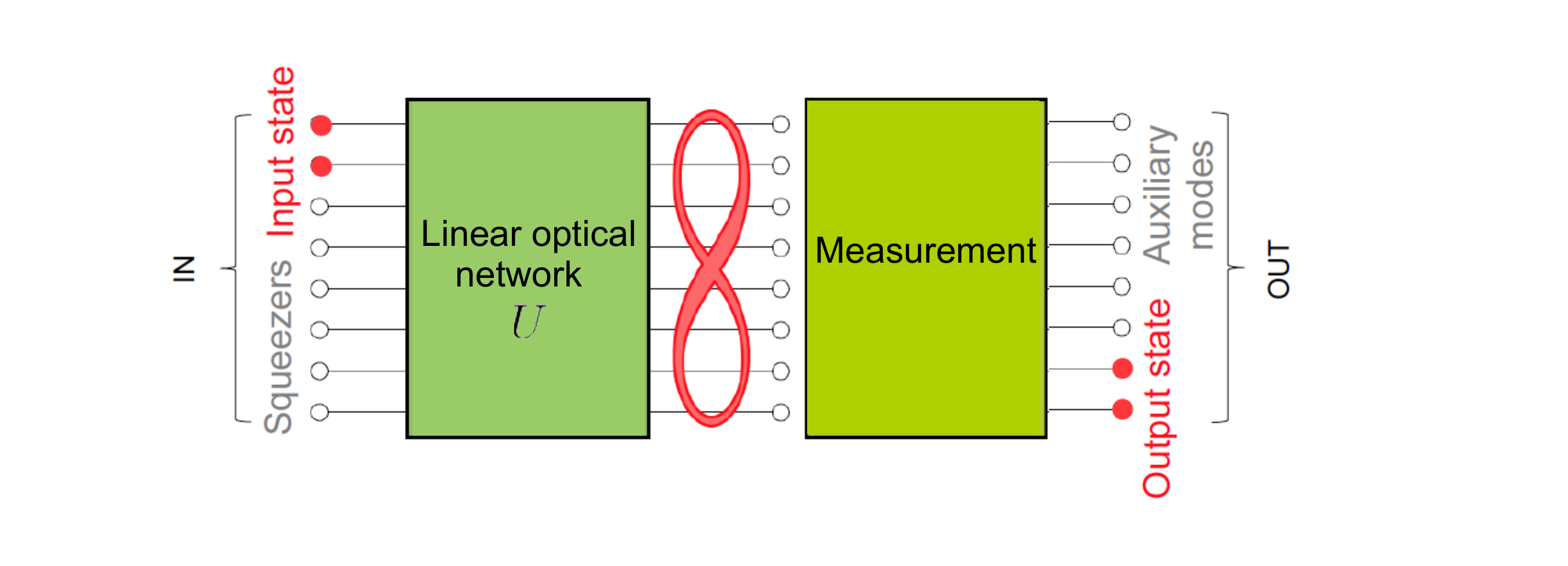}
\caption{General procedure for Gaussian MBQC. Auxiliary modes are entangled to the input modes carrying the state to be processed. After that, suitable measurements are performed on the auxiliary modes, such that the remaining (un-measured) modes are left in a transformed output state. In the figure, for consistency with the notations used in this paper and in particular with Eq.(\ref{eqtrasf-y0}), we have incorporated the choice of the homodyne measurement angles in the definition of the matrix $U$, such that measurement is performed along the same quadrature (e.g., $\hat p$) in all modes.}
\label{fig:scheme-new}
\end{figure}

The general goal of a Gaussian quantum computation is to perform a desired symplectic transformation $\left( \begin{array}{cccccccc}
A_{\text{res}} & B_{\text{res}}  \\
C_{\text{res}} & D_{\text{res}} \\
\end{array} \right)$ on an input quantum state, to which are associated the quadrature operators $\vec{x}^{\, \text{in}} = (\hat{x}^{\text{in}}_1,...,\hat{x}^{\text{in}}_n)^T$ and $\vec{p}^{\, \text{in}} = (\hat{p}^{\text{in}}_1,...,\hat{p}^{\text{in}}_n)^T$. Namely, we want that
\ba
\label{eq:trasf-simpl0}
\left( \begin{array}{cccccccc}
\vec{x}^{\, \text{out}} \\
\vec{p}^{\, \text{out}} \\
\end{array} \right)  = 
\left( \begin{array}{cccccccc}
A_{\text{res}} & B_{\text{res}}  \\
C_{\text{res}} & D_{\text{res}} \\
\end{array} \right) %
\left( \begin{array}{cccccccc}
\vec{x}^{\, \text{in}} \\
\vec{p}^{\, \text{in}} \\
\end{array} \right) 
\ea
where $\vec{x}^{\, \text{out}}$ and $\vec{p}^{\, \text{out}}$ are the quadrature operators associated to the output modes~\cite{dutta1995real, Ukai2010b}. The corresponding annihilation operators are $\vec{a}^{\, \text{in}} = (\hat{a}^{\text{in}}_1,...,\hat{a}^{\text{in}}_n)^T$ and $\vec{a}^{\, \text{out}} = (\hat{a}^{\text{out}}_1,...,\hat{a}^{\text{out}}_n)^T$, related to the quadrature operators in each mode by $\hat a = (\hat x + i \hat p)/2$.  

In a MBQC strategy, this goal is achieved by using as an ancillary resource $m$ independent $\hat p$-squeezed states, to which can be associated the annihilation operators $\vec{a}^{\, \text{squ}} = (\hat{a}^{\text{squ}}_1,...,\hat{a}^{\text{squ}}_m)^T$ and analogous quadrature operators. 
Hence initially we have in total $n + m = N$ optical modes collectively indicated by $\vec{a}^{\, \text{IN}} = (\vec{a}^{\text{in}}, \vec{a}^{\, \text{squ}})^T$.

In order to perform the MBQC, one starts by applying to these input modes a suitable unitary matrix $U$ which depends upon the desired symplectic transformation to implement, corresponding to a linear optical transformation
\be
\label{eqtrasf-y0}
\vec{a}^{\text{OUT}} \, =  U \vec{a}^{\, \text{IN}}.
\ee
This transformation has generally the effect of generating quantum correlations between the modes of the input state and the ancillary modes.
Then it follows  
the measurement of all the system modes (e.g., of the $\hat p$ quadrature) except for the last $n$-ones $\vec{a}^{\, \text{out}} = (\hat{a}^{\text{out}}_1,...,\hat{a}^{\text{out}}_n)^T$, containing the result of the computation \cite{Ukai2010b, ferrini2013compact}. The collective label $\vec{a}^{\text{OUT}}$ indeed stands for $\vec{a}^{\text{OUT}} = (\vec{a}^{\text{aux}}, \vec{a}^{\, \text{out}})^T$ where the output auxiliary modes are $\vec{a}^{\, \text{aux}} = (\hat{a}^{\text{aux}}_1,...,\hat{a}^{\text{aux}}_m)^T$ (see Fig.\ref{fig:scheme-new} where the output auxiliary modes are represented, as well as the relevant output and input modes). All the measurements may be performed simultaneously without harming the determinism of the operation, as for Gaussian operations adaptivity is trivial and can be taken into account by classical corrections in the post-processing stage~\cite{Nielsen2006,Gu2009}. 

Expressing the linear system of Eq.(\ref{eqtrasf-y0}) in the quadrature representation, it is possible to isolate the anti-squeezed quadratures $\hat{x}^{\text{squ}}_i$ for $i = 1, ..., m$ as a function of the squeezed quadratures $\hat{p}^{\text{squ}}_i$, the $m$  individual measurement results $p^{\text{aux}}_i$, as well as the input modes quadratures $\vec{p}^{\text{in}}$ and $\vec{x}^{\text{in}}$. These expressions for $\hat{x}^{\text{squ}}_i$ can be replaced in the expression for the  output modes (i.e., the unmeasured modes) quadratures, which as a consequence become a function of $\vec{p}^{\text{in}}$, $\vec{x}^{\text{in}}$, $p^{\text{aux}}_i$, and $\hat{p}^{\text{squ}}_i$. In Appendix \ref{eq:app-mbqc} we provide an explicit derivation in the particular case of a single-mode input state. The general result reads 
\ba
\label{eq:trasf-simpl-eff}
\hspace{-0.3cm}
\label{eq:general-op}
\left( \begin{array}{cccccccc}
\vec{x}^{\, \text{out}} \\
\vec{p}^{\, \text{out}} \\
\end{array} \right)  = 
\left( \begin{array}{cccccccc}
A & B  \\
C & D \\
\end{array} \right) %
\left( \begin{array}{cccccccc}
\vec{x}^{\, \text{in}} \\
\vec{p}^{\, \text{in}} \\
\end{array} \right) + 
\left( \begin{array}{cccccccc}
\vec{\delta}_x \\
\vec{\delta}_p \\
\end{array} \right) +
\left( \begin{array}{cccccccc}
\vec{\eta}_x \\
\vec{\eta}_p \\
\end{array} \right)
\ea
where ${{\delta}_x}^i = \sum_{j = 1}^{m} c^{i j}_x \hat{p}^{\text{squ}}_j $ and ${{\delta}_p}^i = \sum_{j = 1}^{m} c^{i j}_p \hat{p}^{\text{squ}}_j $
are the undesired noise operators due to finite squeezing in the ancillary input states while ${\eta}_{x}^i  = \sum_{j = 1}^{m} l^{i j}_x {p}^{\text{aux}}_i $, ${\eta}_{p}^i  = \sum_{j = 1}^{m} l^{i j}_p {p}^{\text{aux}}_i$ are real numbers, linear functions of the measurement outcomes. The latter do not affect the symplectic structure of the input-output transformation~\cite{Nielsen2006,Gu2009}; they can be eliminated by classical correction, either implemented optically with feedback, or by digital post-processing if all the modes are to be measured.
The matrices $A$, $B$, $C$, $D$, $c^{i j}_{x, p}$ and $\eta^{i j}_{x, p}$ depend upon the specific transformation $U$ that is applied to the input modes according to Eq.(\ref{eqtrasf-y0}).  The output modes of Eq.(\ref{eq:trasf-simpl-eff}) encode the result of the QC, and their measurement  (which can be simultaneous with the others) provides the result.

In the standard MBQC approach, the unitary matrix $U$ in Eq.(\ref{eqtrasf-y0}) results from the product of three matrices: the unitary $U_V$ constructing the cluster of adjacency matrix $V$ from the input squeezed modes, a beamsplitter interaction which couples the input state and $n$ corresponding modes of the cluster, and a  diagonal matrix $D_{\text{meas}}$ specifying each mode's measurement quadrature (conventionally along this article, when not specified otherwise we will assume that the $p$ quadratures of the resulting modes are measured):
\be
\label{eq:q_comp}
U = U_{\text{comp}} = D_{\text{meas}} U_{\text{BS}} U_{V}.
\ee
The excess noise remaining in Eq.(\ref{eq:trasf-simpl-eff}) can be in this case recast in terms of the cluster nullifiers $\hat{\zeta}_i = \hat{p}_i - \sum_{j = 1}^m V_{ij}  \hat{x}_j$~\cite{vanLoock2007,Ukai2010b}. The variance of these combinations of quadraturs goes to zero in the limit of infinite squeezing~\footnote{For a description in terms of non-Hermitian complex-valued nullifiers, which possess zero variance even in the finite squeezed case, see Ref.\cite{menicucci2011graphical}}.

This choice of linear optics transformation $U$ is not the only possible one, and other choices may be more advantageous, corresponding to different experimental configurations. In the approach that we propose, the matrix $U$ is chosen such that the output modes described by (\ref{eq:trasf-simpl-eff}) match the result of the desired quantum computation. Formally, this can be expressed by the fact that the function 
\be
\label{eq:fitness-last}
f_{1} = \| \left( \begin{array}{cccccccc}
A & B  \\
C & D \\
\end{array} \right) - \left( \begin{array}{cccccccc}
A_{\text{res}} & B_{\text{res}}  \\
C_{\text{res}} & D_{\text{res}} \\
\end{array} \right) \|
\ee
should be as small as possible, where $A_{\text{res}}, B_{\text{res}}, C_{\text{res}}, D_{\text{res}}$ are coefficients of the desired resulting operation as expressed by Eq.(\ref{eq:trasf-simpl0}), and the norm is a standard matrix norm, e.g. the Frobenious norm $\| A \| = \sqrt{ \Sigma_{i,j} | A_{i,j}|^2 }$. 
Hence practically, one can optimize $U$ by minimizing $f_1$. 

The standard choice of unitary matrix $U$ in Eq.(\ref{eq:q_comp}) leads to a value of zero for the $f_1$ function of Eq.(\ref{eq:fitness-last}). 
In Sec.\ref{se:DMBQC}  though we will show that a tangible optical network is actually unnecessary for collecting statistics corresponding to detection of each cluster mode's quadrature or the end result of a QC, thereby yielding values of the function $f_1$ that are close to zero. Furthermore, within our approach one could also specifically address, as a simultaneous task of a multi-objective optimization strategy, the reduction the excess noise incurred by finite squeezing by also minimizing~
\footnote{In the protocols that we will be considering, we will suppose that a single quadrature is measured in the output mode. However, we chose the definition of the noise figure of merit Eq.(\ref{eq:fitness3}) because this provides an upper bound to the noise in the associated output modes, that due to trace preservation under change of basis also incorporates possible noise correlations between different modes and/or quadratures.}
\be
\label{eq:fitness3}
\hspace{-0.25cm} f_2 = \sum_{i=1}^n \lt  \Delta^2 \delta_x^i + \Delta^2 \delta_p^i \rt.
\ee  
%

Before turning to the explanation of our direct MBQC method in Sec.\ref{se:DMBQC}, we further motivate the use of ancillary squeezed states for information processing in the forthcoming Sec.\ref{se:charac}.
 

\section{Characterization of the Gaussian operations accessible by linear optics networks, measurement and post-processing}
\label{se:charac}

In the case that one wants to sample the quadratures corresponding to the results of a desired QC, as indicated by Eq.(\ref{eq:trasf-simpl0}), one could aks the question: is it necessary to use ancillary squeezed states as presented in Sec.\ref{se:MBQC}, or could one just measure the input modes in a suitable basis, possibly after having mixed them on a beam-splitter, and allowing for post-processing of the collected statistics, without relying upon the use of ancillary squeezed states?

In this section we determine the most general transformation on the input state that is achievable via the combination of these three tools, namely: 1) A linear optics network 2) Measurement in an arbitrary basis via homodyne detection 3) Post-processing. 

The first two elements 1) and 2) above amount to the following transformation, which contains both phase shifters and a change of basis:
\be
\label{eq:general-op2}
\left( \begin{array}{cccccccc}
\vec{x}^{\,\text{out}} \\
\vec{p}^{\,\text{out}} \\
\end{array} \right)  = 
\left( \begin{array}{cccccccc}
X & - Y  \\
Y & X \\
\end{array} \right) %
\left( \begin{array}{cccccccc}
\vec{x}^{\,\text{in}} \\
\vec{p}^{\,\text{in}} \\
\end{array} \right).
\ee
In order for this matrix to be symplectic orthogonal, i.e. to yield a proper change of basis corresponding to the correct commutation relations in the output, the matrices $X$ and $Y$ must satisfy~\cite{dutta1995real}
\ba
&& X X^T + Y Y^T = \mathcal{I} \label{eq:1} \\
&& X Y^T = Y X^T  \label{eq:2}.
\ea 

Next,  by post-processing we mean digital recombination of the data acquired. Assume, as we do all along this article, that the quadrature $\hat p$ is measured on each mode. One can then recombine, for example, the traces associated with the operators $\hat p_1$ and $\hat p_3$ as $\hat p' = 1/\sqrt{2}(\hat p_1 +\hat p_3)$. This effectively yields a measurement outcome of the quadrature $\hat p$ associated with the mode $\hat a' = 1/\sqrt{2}(\hat a_1 +\hat a_3)$. Note that only real transformations are allowed, because no information about the quadrature $\hat x$ can be obtained once that $\hat p$ has been measured. Similarly, on a single mode one can apply a gain factor to the quadrature measurement, yielding squeezing or dilation.
These two transformations are summarized respectively by the matrices:
\be
\label{eq:post-proc}
\left( \begin{array}{cccccccc}
O &  0 \\
0 & O \\
\end{array} \right);
\left( \begin{array}{cccccccc}
R^{-1} &  0 \\
0 & R \\
\end{array} \right)
\ee
where $O$ is a real orthogonal matrix, i.e. $O O^T = 1$, and $R$ is a positive real diagonal matrix. Matrices of these two kinds can be arbitrarily combined. Let us focus on the $p$-block, as it is the one which is measured. The product will be of the form: $R_1 R_2 O_1 O_2 O_3 R_3 O_4...$. For obvious group properties this can be expressed without loss of generality (redefining the group elements) as $R_1 O_1 R_2 O_2 ...R_h O_h$ with $h$ integer. The result of this combination will always be a real square matrix. As such, it admits a singular value decomposition which allows expressing it as $O R O'$ where $O'$ is another real orthogonal matrix.

With this in mind, the full transformation that can be implemented under the three tools defined above reads
\ba
\label{eq:trivial-opB}
 \hspace{-0.25cm} \left( \begin{array}{cccccccc}
\vec{x}^{\, \text{out}} \\
\vec{p}^{\, \text{out}} \\
\end{array} \right)  
 \hspace{-0.1cm} =  \hspace{-0.1cm} 
\left( \begin{array}{cccccccc}
O &  0 \\
0 & O \\
\end{array} \right)
 \hspace{-0.1cm}
\left( \begin{array}{cccccccc}
R^{-1} &  0 \\
0 & R \\
\end{array} \right)
 \hspace{-0.1cm}
\left( \begin{array}{cccccccc}
X' & - Y'  \\
Y' & X' \\
\end{array} \right) %
 \hspace{-0.1cm}
\left( \begin{array}{cccccccc}
\vec{x}^{\,\text{in}} \\
\vec{p}^{\, \text{in}} \\
\end{array} \right) 
\ea
where the orthogonal matrix $O'$  has been re-absorbed in the first symplectic matrix by exploiting the group property of the unitary group (which is a maximally compact subgroup of the symplectic group), i.e. $X' = O' X$ and $Y' = O' Y$.

This transformation is to be compared to the most general Gaussian transformation (disregarding displacements) described by the evolution $H_G$. The latter can be expressed in the Heisenberg representation by virtue of the Bloch-Messiah decomposition as~\cite{dutta1995real}
\be
\hspace{-0.3cm}
\label{eq:general-op}
\left( \begin{array}{cccccccc}
\vec{x}^{\,\text{out}} \\
\vec{p}^{\,\text{out}} \\
\end{array} \right)  = 
\left( \begin{array}{cccccccc}
\mathcal{X} & - \mathcal{Y}  \\
\mathcal{Y} & \mathcal{X} \\
\end{array} \right) %
\left( \begin{array}{cccccccc}
K^{-\fr{1}{2}} & 0  \\
0 & K^{\fr{1}{2}} \\
\end{array} \right) 
\left( \begin{array}{cccccccc}
\mathcal{X}' & - \mathcal{Y}'  \\
\mathcal{Y}' & \mathcal{X}' \\
\end{array} \right) %
\left( \begin{array}{cccccccc}
\vec{x}^{\,\text{in}} \\
\vec{p}^{\,\text{in}} \\
\end{array} \right)
\ee
where the matrices $\mathcal{X}$, $\mathcal{X}'$, $\mathcal{Y}'$ and $\mathcal{Y}$ satisfy respectively the same conditions required in Eqs.(\ref{eq:1}) and (\ref{eq:2}), while $K$ is a real diagonal matrix with positive elements as $R$.

By comparing Eqs.(\ref{eq:trivial-opB}) and (\ref{eq:general-op}), we see that not all the Gaussian operations can be implemented with the authorized tools: in particular, those which require a Bloch-Messiah decomposition (\ref{eq:general-op}) with $\mathcal{Y} \neq 0$ cannot be implemented by this "trivial" measurement scheme, and hence require further tools to be implemented. 

The simplest example is the two-mode entangling $C_Z$ gate, 
\be
\label{Cz}
\left( \begin{array}{cccccccc}
\vec{x}^{\text{out}} \\
\vec{p}^{\text{out}} \\
\end{array} \right)  = 
\left( \begin{array}{cccccccc}
I & 0  \\
V & I \\
\end{array} \right) %
\left( \begin{array}{cccccccc}
\vec{x}^{\text{in}} \\
\vec{p}^{\text{in}} \\
\end{array} \right)
\ee
with $V = \left( \begin{array}{cccccccc}
0 & 1  \\
1 & 0 \\
\end{array} \right)$
which explicit Bloch-Messiah decomposition is of the form given by Eq.(\ref{eq:general-op}) with $\mathcal{X} = \left(
\begin{array}{cc}
x & 0  \\
 0 & x  \\
\end{array}
\right), \mathcal{Y} = \left(
\begin{array}{cc}
0 & y  \\
 y & 0  \\
\end{array}
\right), K^{-\fr{1}{2}}  = \left(
\begin{array}{cc}
k & 0  \\
 0 & k  \\
\end{array}
\right), \mathcal{X}' = \left(
\begin{array}{cc}
x' & 0  \\
 0 & x'  \\
\end{array}
\right), \mathcal{Y}' = \left(
\begin{array}{cc}
0 & y'  \\
 y' & 0  \\
\end{array}
\right)$ with  $x =  \frac{1+\sqrt{5}}{2 \sqrt{5+2 \sqrt{5}}}$,  $y =  \frac{3+\sqrt{5}}{2 \sqrt{5+2 \sqrt{5}}}$, $k = \frac{1}{2} \left(1+\sqrt{5}\right), x' = \frac{1+\sqrt{5}}{\sqrt{2 \left(5+\sqrt{5}\right)}}, y' =  \frac{1-\sqrt{5}}{\sqrt{10-2 \sqrt{5}}}$. This transformation, instead, can be implemented by MBQC, i.e. making use of ancillary squeezed states, either in the traditional cluster-based approach~\cite{Ukai2010b}, or by the direct approach  that we will detail in the next Section.  In the latter context, the minimal number of ancillary squeezed states to be used is determined in Appendix \ref{se:minimal}.

\section{Directly synthesized cluster states and MBQC}
\label{se:DMBQC}

\sse{Direct approach to MBQC}

\begin{figure}
\centering
\includegraphics*[width=1.05\columnwidth]{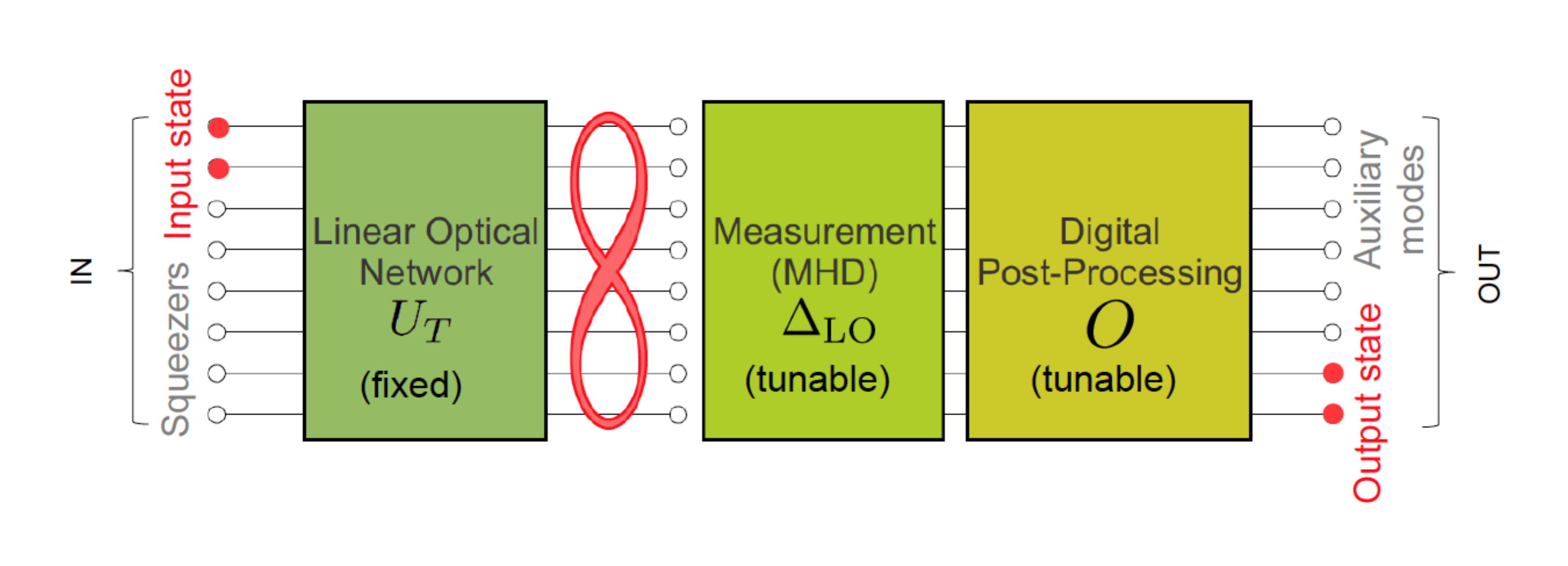}
\caption{Schematic for the construction of quantum operations with a multimode homodyne detector (MHD) followed by digital post-processing.}
\label{fig:scheme}
\end{figure}

\begin{figure} 
\begin{minipage}{\columnwidth}
\includegraphics*[width=\columnwidth]{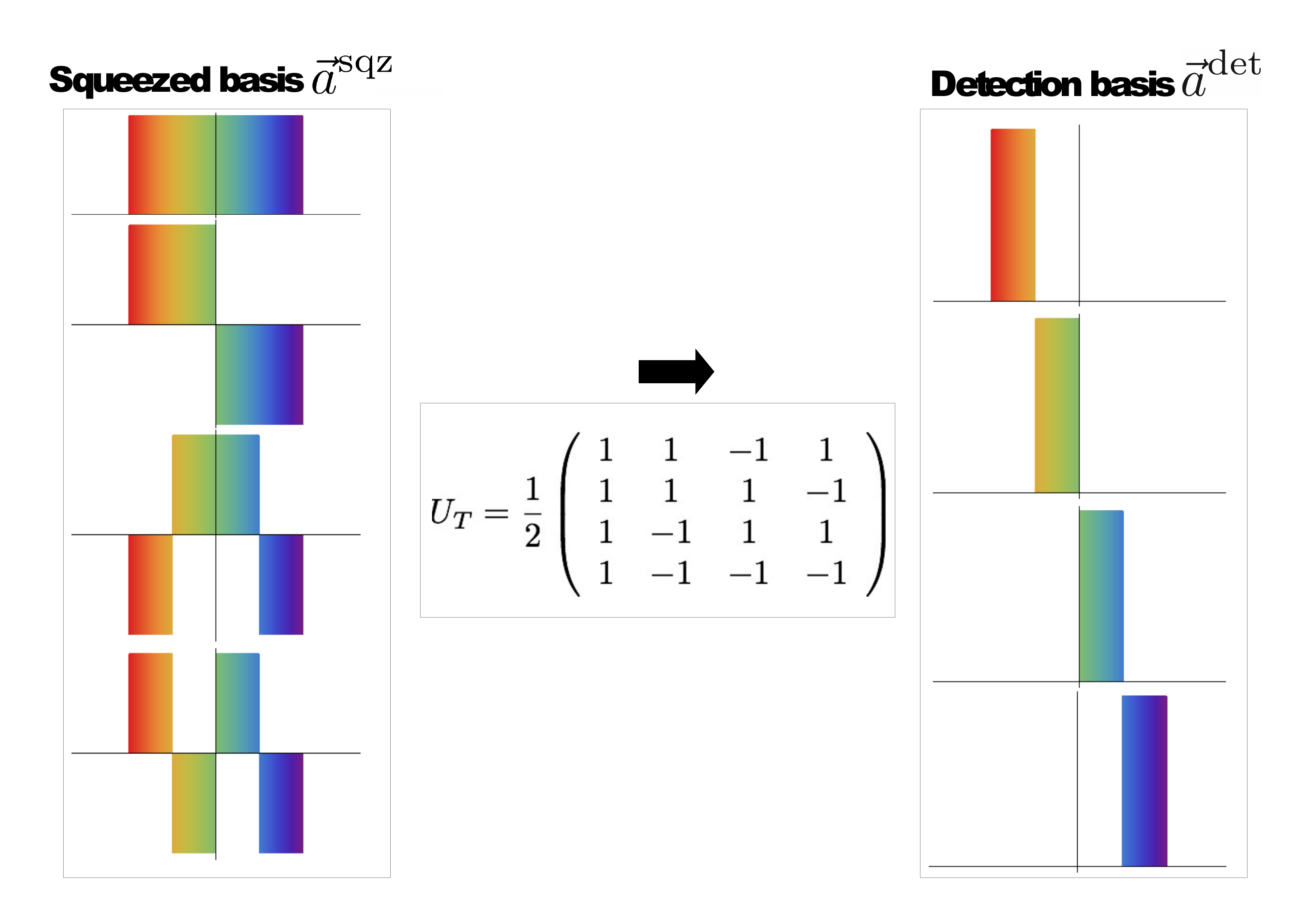}
\end{minipage}
\caption{Basis transformation between the input squeezed modes (left) and the MHD detection modes (right). Left: In the input mode basis, the state is described by a series of independently squeezed states (one or more of the modes may also encode the state to be processed). These can be either spatial modes (in this case the $x$-axes refers to a spatial coordinate), or frequency modes, as e.g. in the experiment of Ref.~\cite{Roslund_13b} (in this case the $x$-axes refers to frequency). Right: The alternate basis of the {\it pixel modes} (either spatial or in the frequency domain) can be chosen to measure the same multi-mode state. The basis change relating these two bases is described by a unitary matrix. In the case of the toy-model here represented, this unitary matrix is particularly simple, and can be guessed by simply looking at how the modes at the right decomposes onto pixels-modes: this provides the column of the matrix, given the definition in Eq.(\ref{eq2}). 
Colors only provide a pictorial guide for the eye, with no quantitative scale. The $y$-axes represents the mode intensity. 
\label{fig:UT}}
\end{figure}

We consider a general experimental scheme in which a set of input modes as well as ancillary $\hat p$ squeezed modes $\vec{a}^{\, \text{IN}} = (\vec{a}^{\text{in}}, \vec{a}^{\, \text{squ}})^T$ 
are interrogated in an alternative basis $\vec{a}^{\, \text{det}}$ by a set of independent homodyne detectors. These detectors implement a multimode homodyne detection (MHD) as seen in Fig.\ref{fig:scheme}. 
Appropriately, the detection modes are viewed as resulting from a linear transformation of the independently squeezed modes~\cite{Braunstein2005-irreducible}:
\be
\label{eq2}
\vec{a}^{\, \text{det}} = {U_T} \vec{a}^{\, \text{IN}},
\ee
i.e. the MHD performs a change of basis. An example of this change of basis is provided in Fig.\ref{fig:UT}. 
Each homodyne detection is implemented on a given quadrature by choosing the phase of the local oscillator in each detection mode, which is modeled by a diagonal matrix $\Delta_{\mathrm{LO}}$ with complex elements of unit modulus. 
Following detection, the acquired homodyne traces are digitally recombined in a post-processing stage, leading finally to a transformation equivalent to the one described in Sec.\ref{se:charac} and defined in Eq.(\ref{eq:trivial-opB}), but acting on many more modes.
As we will show in a moment, we'll be able to implement desired MBQC operations even when the post-processing consists in a real orthogonal matrix $O$ only, i.e. without using a further squeezing matrix $R$. Hence, the total transformation effectuated by the MHD plus post-processing takes the form: 
\ba
\label{eq:trtot_m}
\hspace{-0.2cm} \vec{a}^{\text{OUT}} = O(\vec{\theta}) \Delta_{\mathrm{LO}} (\vec{\varphi}) U_T \, \, \vec{a}^{\, \text{IN}}  \equiv U_{\mathrm{MHD}} (\vec{\theta}, \vec{\varphi})  \, \,  \vec{a}^{\, \text{IN}},
\ea
each step being represented in Fig.\ref{fig:scheme}. Here again, we assume that the $p$ quadrature of the modes is finally obtained. This transformation hence mimics the application of a unitary transformation on the input and ancillary squeezed modes exactly as in Eq.(\ref{eqtrasf-y0}). 

The transformation in Eq.(\ref{eq:trtot_m}) contains tunable degrees of freedom, namely the local oscillator phases $\Delta_{\mathrm{LO}} (\vec{\varphi})$, and the post-processing 
$O (\vec{\theta})$, which may be optimized so as to achieve the desired output $\vec{a}^{\mathrm{OUT}}$. For example, they may be chosen so that $\vec{a}^{\mathrm{OUT}}$ replicates the statistics corresponding to a direct cluster state measurement~\cite{ferrini2013compact}. 
Alternatively, the transformation can be customized such that measurement of $n$ chosen output modes yields the statistics of the readout mode following a desired QC on an input state.
It is worth noting that this method subsumes creation of the QC resource state into the state measurement itself, which reduces the quantum depth to a value of one for the ensemble of these two stages~\cite{Elham2007b,Browne2007}.

Importantly, a \emph{post-facto} examination of arbitrary linear combinations of the collected data is entirely equivalent to a direct optical transformation of the modes according to Eq.~(\ref{eq:trtot_m}) followed by their detection. 
This equivalence is due to the fact that the matrix $O$ is real orthogonal and does not mix the field quadratures (i.e., $\vec{q}^{\mathrm{\, OUT}}$ commutes with $O^{-1} \vec{q}^{\mathrm{\, OUT}}$, and $\vec{p}^{\mathrm{\, OUT}}$ with $O^{-1} \vec{p}^{\mathrm{\, OUT}}$).

This strategy is applicable for quantum computation.
Toward that end, 
the primary conceptual advance taken in this work envisions the measurement of a final mode's state  - the QC result -
as an outcome of Eq.(\ref{eq:trasf-simpl-eff}) with $U=U_{\mathrm{MHD}}$ in Eq.(\ref{eqtrasf-y0}). Consequently, it is possible to make no formal assumptions as to the structure of $U_{\mathrm{MHD}}$ and instead directly minimize Eq.(\ref{eq:fitness-last}) (and possibly (\ref{eq:fitness3})) on 
the free parameters; in particular, $U_{\mathrm{MHD}}$ may not be related to any unit-weight cluster state matrix.
The emphasis on the measurement outcome, rather than the building block operations necessary to achieve it, represents a new approach, which we shall refer to as ``direct MBQC".
It is important to stress the conceptual difference between this approach and that of Ref.~\cite{ferrini2013compact}. Namely, effort was directed in \cite{ferrini2013compact} toward selecting a $U_\text{MHD}$ that matches a target $U_{\text{comp}}$. In contrast, the present work is outcome-oriented and takes $U_\text{MHD}$ to be that which minimizes Eq.(\ref{eq:fitness-last}) with no concern for its specific structure. 



\sse{A simple example: Fourier transform on a single-mode state}
\label{sse:FT}

In order to exemplify our method, it is useful to start by presenting a simple single-mode transformation, that was already used in Refs.~\cite{Furusawa2011, ferrini2013compact, Ferrini-prep} to illustrate ideas related to MBQC protocols, and that is part of the elementary gates set for universal quantum computing~\cite{Gu2009}, i.e. the Fourier transformation $\left(  \begin{array}{cccccccc} 
\hat{x}^{\text{out}} \\
\hat{p}^{\text{out}} \\ \end{array}
\right) =
\left(  \begin{array}{cccccccc}
0 & -1 \\
1 & 0 \\
\end{array} \right)
\left( \begin{array}{cccccccc}
\hat x^{\text{in}} \\
\hat p^{\text{in}} \\
\end{array} \right) =
\left(  \begin{array}{cccccccc}
- \hat p^{\text{in}} \\
\hat x^{\text{in}} \\
\end{array} \right)$. 
For this simple example a trivial solution of the kind presented in Sec.\ref{se:charac} exists. 
In the standard MBQC approach, the proper measurement matrix $D_{\text{meas}}$ along with the $U_{V}$ necessary to implement this QC by a three-mode cluster state are the ones reported in Ref.~\cite{ferrini2013compact} (see also the Appendix of Ref.\cite{Ferrini-prep}). The calculation of the output mode containing the computation result, presented in the same Appendix, follows the lines of Refs.~\cite{vanLoock2007,Furusawa_Book} and yields
\ba
\label{eq:teleport_result4}
\hat{x}^{\text{out}}  &=&  -\hat{p}^{\text{in}} + {p}^{\text{aux}}_3 - \sqrt{2} {p}^{\text{aux}}_2 - \hat{\zeta}_2  \\
\hat{p}^{\text{out}} &=&   \hat{x}^{\text{in}}   - \sqrt{2} p^{\text{aux}}_1  - \hat{\zeta}_1 + \hat{\zeta}_3  \nn,
\ea
where $\hat{\zeta}_{i}$ are the previously defined nullifiers. 

We consider input modes whose squeezing levels correspond to those seen in the four-mode multimode state of Ref.~\cite{Roslund_13b}.  Specifically, the squeezed quadrature variances relative to shot noise (SN) were in that case $-7$dB, $-6$dB, $-4$dB, and $0$dB. Here  the fourth (minimally squeezed) mode serves as the input mode. Taking the expression of the linear optics matrix building the three-mode cluster state $U_{V}$ as given in Ref.~\cite{ferrini2013compact}, the excess noise quadratures $\Delta^2 {\delta}_x = \Delta^2 \hat{\zeta}_2 $ and $\Delta^2 {\delta}_p = \Delta^2(- \hat{\zeta}_1 + \hat{\zeta}_3 )$ in Eq.(\ref{eq:teleport_result4}) are detailed in Table~\ref{table-QC} 
\footnote{As opposed to the output modes of Eq.(\ref{eq:teleport_result2}), the extra noise modes are not constrained by the uncertainty principle $\Delta^2 {\hat{x} } \Delta^2 {\hat{p }} \geq 1$ as they are added contributions, and can display $\Delta^2  {\delta}_x +\Delta^2  {\delta}_p  = 0$ in the limit of high squeezing. The SN limit yields $\Delta^2  {\delta}_x = 3$ and $\Delta^2 {\delta}_p = 2$, i.e. $\Delta^2  {\delta}_x + \Delta^2 {\delta}_p= 5$.}. 

\begin{table}
\centering
\begin{tabular}{ l | c |  c |   c | r | } 
                  
   {} & $\Delta^2  {\delta}_x$ & $\Delta^2  {\delta}_p $ & $f_1$    & $f_2$   \\
    \hline        
  $U_{\text{comp}}$ from Ref.\cite{ferrini2013compact}   & 1.20 &  0.48 & 0 &   1.68 \\   \hline  
  Optimized matrix $U_\text{MHD}^{\text{best}}$ & 0.25 &  0.18 & $10^{-16}$ & 0.43 \\ \hline

\end{tabular}
\caption{Comparison of the QC's excess noise and approximation of the result in the standard and direct method, relative to the implementation of a Fourier transform on a single-mode input state.}
\label{table-QC}
\end{table}

Let us now turn to our direct approach.  Selecting the Fourier transform as the desired operation to implement dictates that the coefficients of Eq.(\ref{eq:trasf-simpl0}) are real numbers and must be taken as $A_{\text{res}}=0$, $B_{\text{res}} = -1$, $C_{\text{res}} = 1$, $D_{\text{res}} = 0$. 
For this example, the squeezed mode basis, the detection basis, and the transformation $U_T$ specified in Eq.(\ref{eq2}) are shown in Fig.\ref{fig:UT} (modulo the mode re-labeling above mentioned). This detection basis is similar to that of Ref.~\cite{Roslund2013, Roslund_13b} in which squeezed Hermite-Gauss modes in the frequency domain are detected in a basis consisting of slices of the spectrum. 
A minimization of Eq.(\ref{eq:fitness-last}), which takes the squeezing levels already considered, yields the output mode: 
\ba 
\label{eq:teleport_result2} \hspace{-0.23cm} \hat{x}^{\text{out}} \hspace{-0.03cm} &=& \hspace{-0.03cm} -1. \,  \hat{p}^{\text{in}} + 0.2 \, \hat{p}_1^{\text{squ}}-0.97  \hat{p}_2^{\text{squ}} -0.14 \hat{p}_3^{\text{squ}}  \\
& &   -0.82 {p}^{\text{aux}}_2 -0.57{p}^{\text{aux}}_3 -0.03{p}^{\text{aux}}_1  \nn \\
\hspace{-0.1cm} \hat{p}^{\text{out}}  \hspace{-0.03cm} &=&  \hspace{-0.03cm} 1. \,  \hat{x}^{\text{in}} -0.4  \hat{p}_1^{\text{squ}} -0.34   \hat{p}_2^{\text{squ}} +0.54  \hat{p}_3^{\text{squ}}  \\
& &+0.10  {p}^{\text{aux}}_2 -0.47 {p}^{\text{aux}}_3 +0.58 {p}^{\text{aux}}_1. \nn 
\ea
More details concerning the numerical procedure can be found in Appendix \ref{se:how-optimization-is-performed}.
Note that an outcome of this measurement provides only the $\hat{p}$ quadrature. On doing so, we find the values of $\Delta^2  {\delta}_x$ and  $\Delta^2  {\delta}_p$ reported in Table~\ref{table-QC},
which corresponds to a $\sim 74 \%$ reduction in the excess noise as compared to that arising from application of $U_{\text{comp}}$. This reduction of noise relative to the traditional approach comes at the expense of having, in principle, an approximate solution; however, in the considered example, the solution is practically exact, i.e. it exhibits an error on the order of the numerical precision of the machine used for the optimization (one part in $10^{16}$). Incidentally, we remark that relatively to this simple example the direct method outperforms, in terms of the added noise figure of merit, the standard MBQC method even when the latter uses - as we have addressed in Ref.~\cite{Ferrini-prep} - a cluster state constructed with optimal linear optics network: that method resulted indeed in an intermediate noise reduction corresponding to $f_2 = 0.43$.

It is interesting to consider whether the structure of the optimal $U_\text{MHD}$ reproduces that of a particular $U_{\text{comp}}$ in Eq.(\ref{eq:q_comp}) (i.e.,  it may be decomposed in terms of a teleportation onto a cluster by a beam-splitter interaction, followed by successive measurements). The matrix distances between $U_\text{MHD}^{\text{best}}$ and a series of potential $U_{\text{comp}}$ are examined; however, 
it does not prove feasible to discover a $U_{\text{comp}}$ that approaches $U_\text{MHD}^{\text{best}}$
\footnote{The distance $\text{Norm}_F \lqu U_V (\vec{\theta})  - U_{\text{MHD}}^{\text{best}} \rqu$ is used as a metric of matrix similarity,  where $ U_V (\vec{\theta})$ are all the possible transformations leading to the same cluster~\cite{Ferrini-prep}. Even upon using a free readout angle ${D^{(4,4)}_{\text{meas}}}$, we find a minimal norm $\text{Norm}_F \simeq 2.01$. 
For comparison, when $U_{\text{MHD}}$ matches a given matrix, $\text{Norm}_F$ amounts to the machine numerical error ($\simeq 10^{-16}$).}.
Consequently, the discovered $U_\text{MHD}$ can not be interpreted as a traditional MBQC based on the use of unweighted cluster states. It is a heavy numerical task to address the question whether weighted clusters participate into the teleportation of the input state across the ancillary modes. In this paper we precisely want to stress the possibility of pragmatically optimize the experimental parameters to achieve a given quantum computation, disregarding its interpretation in terms of a cluster state.

\sse{A more relevant example: implementation of a  $C_Z$ by means of the direct approach}

Let us now consider an arbitrary two-mode input state.
As we have seen in Sec.\ref{se:charac}, the  $C_Z$ gate in Eq.(\ref{Cz}) does not belong to the class of operations that can be implemented by measurement and post-processing of the input state modes only. 
Ancillary squeezed states are needed instead.  In the traditional MBQC setting, a four-mode cluster state is used to implement this gate, resulting in a total of six optical modes~\cite{Nielsen2006}. We refer this time to a six-mode run of the experiment reported in Refs.~\cite{Roslund2013, Roslund_13b} as a prototypical situation.
In that experiment, the fixed change of basis $U_T$ reported in Appendix \ref{se:exp-values}  is implemented by detecting the available squeezed modes in the 6-mode pixel basis. 
We numerically minimize the function $f_1$ in Eq.(\ref{eq:fitness-last}) over the free tunable parameters $O$ and $\Delta_{\text{LO}}$.
This results in the output noise operators

\ba
\label{eq:extra-noise-terms-cz}
{\delta_x}_1 &=& 0.16  \hat{p}^{\text{squ}}_1 + 0.74  \hat{p}^{\text{squ}}_2 + 0.23 \hat{p}^{\text{squ}}_3 - 0.46 \hat{p}^{\text{squ}}_4 \nn \\
{\delta_x}_2 &=& -0.32 \hat{p}^{\text{squ}}_1  + 0.38 \hat{p}^{\text{squ}}_2 - 0.76 \hat{p}^{\text{squ}}_3 + 0.31 \hat{p}^{\text{squ}}_4  \nn \\
{\delta_p}_1 &=& -0.16 \hat{p}^{\text{squ}}_1 - 0.48 \hat{p}^{\text{squ}}_2 + 0.49 \hat{p}^{\text{squ}}_3 - 0.31 \hat{p}^{\text{squ}}_4  \nn \\
{\delta_p}_2 &=& -0.96 \hat{p}^{\text{squ}}_1 - 0.53 \hat{p}^{\text{squ}}_2 + 0.23 \hat{p}^{\text{squ}}_3 - 0.13 \hat{p}^{\text{squ}}_4   \nn
\ea
and measurement outcomes
\ba
{\eta_x}_1 &=& 0.21 {p}^{\text{aux}}_1 + 0.63 {p}^{\text{aux}}_2 + 0.49 {p}^{\text{aux}}_3 - 0.39 {p}^{\text{aux}}_4  \nn \\
{\eta_x}_2 &=& 0.64 {p}^{\text{aux}}_1 - 0.37 {p}^{\text{aux}}_2 + 0.39 {p}^{\text{aux}}_3 + 0.47 {p}^{\text{aux}}_4  \nn \\
{\eta_p}_1 &=&  0.76 {p}^{\text{aux}}_1 + 0.44 {p}^{\text{aux}}_2 - 0.61 {p}^{\text{aux}}_3 + 0.67 {p}^{\text{aux}}_4  \nn \\
{\eta_p}_2 &=&  1. {p}^{\text{aux}}_1 + 0.33 {p}^{\text{aux}}_2 - 0.47 {p}^{\text{aux}}_3 - 0.95 {p}^{\text{aux}}_4.  \nn
\ea
Even in this case, the fitness function $f_1$ obtained is of the order of the internal precision of the machine used for the optimization, i.e. a practically exact solution is found which yields the 
output state Eq.(\ref{eq:trasf-simpl-eff}) with symplectic matrix specified by Eq.(\ref{Cz}).
Furthermore, we again observe a conspicuous reduction in the extra noise associated with the output modes (67 \%), as compared to the standard implementation via cluster state-based MBQC~\cite{Ukai2011exp}. The comparison is reported in Table \ref{table-QC2}.

\begin{table}
\centering
\begin{tabular}{ l | c |  c | c |c |  c | r | } 
                  
   {} & $\Delta^2  {\delta}_{x}^1$ & $\Delta^2  {\delta}_{p}^1 $ & $\Delta^2  {\delta}_{x}^2$ & $\Delta^2  {\delta}_{p}^2 $ & $f_1$    & $f_2$   \\
    \hline        
  $U_{\text{comp}}$ from Ref.\cite{Ukai2011exp}  & 0.41 & 2.35 & 2.79 & 1.32 &   0    &  6.87 \\   \hline  
  Optimized matrix $U_\text{MHD}^{\text{best}}$   & 0.77 & 0.23 & 0.28 & 0.95 & $10^{-15}$ & 2.24  \\ \hline

\end{tabular}
\caption{Comparison of the QC's excess noise and approximation of the result in the standard and direct method relative to the implementation of the $C_Z$ gate on a two-mode input state. We have assumed the following squeezing in the input squeezed modes appearing in Eq.(\ref{eq:extra-noise-terms-cz}): $p_1 = e^{-r_1} p_{0}, p_2 = e^{-r_2} p_{0}, p_3 = e^{-r_3} p_{0}, p_3 = e^{-r_3} p_{0}$ with $r_1 = 0.79, r_2 = 0.36, r_3 = 0.14, r_4 = 0.05$, where $ p_{0}$ is associated to the vacuum.} 
\label{table-QC2}
\end{table}

\sse{Comments on Gaussian universality}

Characterizing the class of the Gaussian operations that can be implemented with our method is not an easy task. Indeed, this strongly depends on the propagation network $U_T$ that implements the change of basis between the squeezed (and input) modes and the detection modes, as well as on the number of ancillary squeezed states. The matrix $U_T$, in turn, depends on the actual experimental implementation. Some relevant "extremal" cases can be however addressed. If, for instance, $U_T$ coincides with the matrix that forms a resource cluster state (i.e., a linear one for single-mode operations, or a square one for multi-mode operations, in sufficiently high dimension) one recovers universal Gaussian operations. When instead the matrix  $U_T$ is the identity, no actual quantum correlation is established between the input and squeezed states. Hence, in this case the ancillary squeezed states are effectively not used, analogously as for the trivial operations discussed in Sec.\ref{se:charac}, a part from post-processing, which however in this case only allows mixing the results of independent measurements of input and squeezed states. 
The case of $U_T$s that we considered in this work, and that are inspired by the experiments at LKB, are intermediate between these two possibilities: correlations between the squeezed and input modes are established by these transformations, but the resulting state is not necessarily a unit-weight cluster state. Yet, some relevant operations can still be performed, as we have shown. 

Establishing the set of Gaussian operations that, given a fixed $U_T$, can be implemented is not a straightforward task either. Cleary, this set includes all the computations that are identified by the necessary and sufficient condition in Ref.\cite{ferrini2013compact}.
However, the present approach allows to address a broader set of computations: some computations may not satisfy Eq.(\ref{eq:q_comp}), and yet could be implementable in our new direct approach. 

In Appendix, we compare the number of degrees of freedom that corresponds to operating an arbitrary symplectic transformation on $n$ modes, to the number of degrees of freedom available with our method, given a certain number of available ancillary squeezed states. This yields a lower bound on the number of ancillary modes that should be used if one wants to implement all the symplectic operations in a given dimension. 

\sse{Targeting cluster states with the direct method}

\ssse{Cluster states}

The approach described may be employed as well  to collect a statistics corresponding to the measurement of a fixed quadrature on all the modes of a certain cluster state. As an example, we consider the fabrication of a 4-mode linear cluster state. 
The matrices $O$ 
and $\Delta_{\text{LO}}$ are now chosen to minimize the function 
\be
\label{eq:f1}
f_3 = \fr{1}{m} \sum_{i = 1}^m \Delta^2 \zeta_i
\ee
where $\hat{\zeta}_i = \hat{p}_i - \sum_{j = 1}^m V_{ij}  \hat{q}_j$ are the cluster nullifiers. 
Referring again to the squeezing distribution and the mode structure of Sec.\ref{sse:FT}, the resultant nullifiers are those reported in Table~\ref{table-cluster}. Each value corresponds to field fluctuations below the SN limit, which indicates successful creation of the cluster state. 
We stress that the nullifiers of this state can not be directly assessed with a single choice of $O$ and $\Delta_{\text{LO}}$ since the phase degree of freedom $\Delta_{\text{LO}}$ has been exploited to create the state itself. However, a common quadrature of all cluster nodes may be measured instead.
\begin{table}

\centering

\begin{tabular}{ l | c | r | } 
                  
  {} & Nullifier variances $\{ \fr{\Delta^2 \zeta_i}{{\Delta^2 \zeta_i}_0} \}$ & $f_3$ \\   \hline     
  Matrix $U_V$ from Ref.~\cite{vanLoock2008}  &  $\{0.20,0.50,0.24,1.0\}$ &  1.16 \\   \hline     
  Optimized matrix $U_\text{MHD}^{\text{best}}$ & \{0.23,0.48,0.21,0.70\} & 0.97 \\   
  \hline

\end{tabular}
\caption{Comparison of nullifier variances for a 4-node linear cluster state with the unitary transformation $U$ and with the direct method.  ${\Delta^2 \zeta_i}_0$ are the shot noise (SN) levels, which are defined as the nullifier variances for vacua inputs.}

\label{table-cluster}

\end{table}

\ssse{Other applications of post-processing}

The digital post-processing currently proposed may be useful in a variety of experimental situations; data acquired from multiple homodyne devices may be analyzed in a manner that reveals information regarding specific mode combinations as if those combinations had been directly measured. For instance, in the experiment of Ref.~\cite{vanLoock2008}, it is possible to reveal multiple clusters with a single optical design and the appropriate post-processing. 
Specifically, the matrix $U_T$ is taken as the usual transformation converting squeezed inputs into a four-node linear cluster state (Eq.(2) of Ref.~\cite{vanLoock2008}). 
Taken alone, this unitary creates the linear cluster as in the original study. However, with an optimal choice of $O$ and $\Delta_{\text{LO}}$, it is also possible to construct a T-cluster $\{ \fr{\Delta^2 \delta_i}{{\Delta^2 \delta_i}_0} \} = \{0.26,0.27,0.27,0.28\}$ and a square cluster $\{0.25,0.25,0.26,0.27\}$ ~\footnote{For the linear cluster we obtain $\{0.57,0.25,0.25,0.55\}$, which is better than in the original experiment as we have taken a pure state realization. The squeezing values are chosen as the minimum ($5.5$dB) and maximum ($6.3$dB) levels provided in Ref.~\cite{vanLoock2008} with a linear interpolation for the remaining two.}.

To further stress that our approach may be employed in the context of other experiments, we note that a protocol to evidence cluster states by exploiting tuning of the homodyne detection phases and post-processing has been considered with cascaded four-wave mixing processes in atomic vapors~\cite{Cai2014}, though to match the obtained transformation with the unitary transformation yielding a cluster state in the spirit of Ref.~\cite{ferrini2013compact}. That system could however with no difficulty be used to implement the direct MBQC protocol introduced here as well.

\ssse{Important remarks}

The power of this software-based method lies in its versatility and reconfigurability. A variety of clusters or QCs may be addressed by only updating the composition of $O$ and $\Delta_{\text{LO}}$, as opposed to a hardware reorganization of the underlying photonic architecture. Conversely, the interest in constructing a traditional quantum network without the inclusion of supplemental post-processing is that measurements of the resultant cluster may be implemented in any quadrature. A limitation of the software approach is indeed the necessity to update the optimized mode transformation for every variation of the detected quadrature.  Nonetheless, a global scan of the local oscillator phase enables accessing both quadratures of cluster modes. 
Reconstruction of a full cluster state covariance matrix, or of a multi-mode state result of a QC, would require multiple optimizations. 

Note that if one wants to \emph{prepare} a quantum mode in the result of a given QC over an input state instead of \emph{sample}  it, one should not rely on post-processing, i.e. $O = \mathcal{I}$, which brings back to $U_{\text{MHD}} = U_{\text{comp}}$ as in the standard MBQC approach - modulo local rotations of the cluster mode and consequent redefinitions of the measurement angles.

\section{Conclusions}
\label{se:concl}

In summary, an original approach to Gaussian MBQC was proposed that does not explicitly rely on the use of cluster states. 
In this method, targeting the desired result of a QC and reducing the associated error due to finite squeezing are achieved by directly tuning the accessible degrees of freedom related to the detection of the resource modes. 
This strategy is readily implementable in several experimental groups, and opens the way for increasingly compact MBQC protocols. In particular, the fact that building an actual cluster state by means of a network of optical elements is un-necessary within our method renders its application possible even in experiments where it may be hard to separate the various squeezed modes, e.g. when these are all propagating in the same optical beam~\cite{Roslund2013}.
  
We stress that our method is especially relevant in cases where one wishes to apply a last Gaussian operation after a non-Gaussian state has been prepared, possibly as an intermediate output of a previous quantum computation. Our scheme indeed remains within the domain of Gaussian transformations, and operating it on a non-Gaussian input state can already allow to solve classically intractable sampling problems~\cite{Mari2012, Rahimi-Keshari2015}.

Implementation of some non-Gaussian operations may also be achievable as a straightforward extension of our method. Indeed, one could use as an ancillary input state not only squeezed states, but also previously prepared non-Gaussian states, such as the cubic phase state $e^{i \hat{q}^3 s} |0 \rangle$. This state, suitably mixed to the input states, may allow implementing non-Gaussian gates such as for instance the cubic phase gate~\cite{Gu2009}. However, in order to deterministically implement further non-Gaussian operations, one would also need to adapt the prepared non-Gaussian states on further modes $e^{i \hat{q}^3 s'} |0 \rangle$ depending on the obtained measurement results. In other words, the deterministic sequential implementation of non-Gaussian operations realized by means of ancillary non-Gaussian input states also requires feed-forward, analogoulsy as to the standard MBQC implementation of non-Gaussian operations. 
As such, feed-forward will prove necessary in order to provide universal quantum computation. 

\section{Acknowledgmenents}

We kindly thank P. van Loock for a careful reading of this manuscript.
This work is supported by the European Research Council starting grant Frecquam, the European Union's (EU) Horizon 2020 research and innovation programme under Grant Agreement No. 665148 and the French National Research Agency project COMB. C. F. and N. T.  are members of the Institut Universitaire de France. J. R. acknowledges support from the European Union through Marie Sklodowska Curie Actions, Y. C. recognizes the China Scholarship Council, and G.F. acknowledges support from the European Union through the Marie Sklodowska-Curie grant agreement No 704192.

\appendix

\section{Generalized formulation for MBQC: explicit procedure for the single-mode case}
\label{eq:app-mbqc}

Consider the situation in which a single-mode input state is to be processed via the use of ancillary independent squeezed states. To the collective operators vector $(\hat{a}_{\text{in}}, \vec{a}^{\text{squ}})$ we apply the general unitary transformation provided in Eq.(\ref{eqtrasf-y0}). 
In order to achieve the most-general single-mode symplectic operation that can be performed on a single-mode input state, 4 ancillary squeezed modes are in principle needed (in the traditional cluster based approach). However, for the most part of single mode operations 3 ancilla modes are sufficient \cite{Ukai2010b} (this is the case for the example of the Fourier transform considered in the main text), and we stick here to the three mode case for the presentation of our  strategy. The generalization to the 4-mode case is straightforward. Expliciting the vectorial structure of Eq.(\ref{eqtrasf-y0}) gives 
\be
\label{eq:boh1}
\left( \begin{array}{cccccccc}
\hat{a}^{\text{aux}}_1 \\
\hat{a}^{\text{aux}}_2 \\
\hat{a}^{\text{aux}}_3 \\
\hat{a}^{\text{out}} \\
\end{array} \right) =
U \left( \begin{array}{cccccccc}
\hat{a}_{\text{in}} \\
\hat{a}_1^{\text{squ}} \\
\hat{a}_2^{\text{squ}} \\
\hat{a}_3^{\text{squ}} \\
\end{array} \right).
\ee
Then in the quadrature representation we can write
\ba
\hspace{-0.8cm}
\label{eq:modi_sys1}
\left\{ \begin{array}{cccccccc}
\hat{x}^{\text{aux}}_1  \hspace{-0.1cm} &=& \hspace{-0.1cm} \hat{x}^{\text{aux}}_1 (\hat x^{\text{in}}, \hat{x}_1^\text{squ}, \hat{x}_2^\text{squ} , \hat{x}_3^\text{squ}, \hat p^{\text{in}}, \hat{p}_1^\text{squ},  \hat{p}_2^\text{squ}, \hat{p}_4^\text{squ})   \\
\hat{x}^{\text{aux}}_2 \hspace{-0.1cm} &=& \hspace{-0.1cm}  \hat{x}^{\text{aux}}_2 (\hat x^{\text{in}}, \hat{x}_1^\text{squ}, \hat{x}_2^\text{squ} , \hat{x}_3^\text{squ}, \hat p^{\text{in}}, \hat{p}_1^\text{squ},  \hat{p}_2^\text{squ}, \hat{p}_4^\text{squ})   \\
\hat{x}^{\text{aux}}_3 \hspace{-0.1cm} &=& \hspace{-0.1cm}  \hat{x}^{\text{aux}}_3 (\hat x^{\text{in}}, \hat{x}_1^\text{squ}, \hat{x}_2^\text{squ} , \hat{x}_3^\text{squ}, \hat p^{\text{in}}, \hat{p}_1^\text{squ},  \hat{p}_2^\text{squ}, \hat{p}_4^\text{squ})   \\
\hat{x}^{\text{out}} \hspace{-0.1cm} &=& \hspace{-0.1cm} \hat{x}^{\text{out}} (\hat x^{\text{in}}, \hat{x}_1^\text{squ}, \hat{x}_2^\text{squ} , \hat{x}_3^\text{squ}, \hat p^{\text{in}}, \hat{p}_1^\text{squ},  \hat{p}_2^\text{squ}, \hat{p}_4^\text{squ})  
\end{array} \right.
\ea
and
\ba %
\hspace{-0.8cm}
\label{eq:modi_sys2}
\left\{ \begin{array}{cccccccc}
\hat{p}^{\text{aux}}_1 \hspace{-0.1cm} &=& \hspace{-0.1cm}\hat{p}^{\text{aux}}_1 (\hat x^{\text{in}}, \hat{x}_1^\text{squ}, \hat{x}_2^\text{squ} , \hat{x}_3^\text{squ}, \hat p^{\text{in}}, \hat{p}_1^\text{squ},  \hat{p}_2^\text{squ}, \hat{p}_4^\text{squ})   \\
\hat{p}^{\text{aux}}_2 \hspace{-0.1cm} &=& \hspace{-0.1cm}  \hat{p}^{\text{aux}}_2 (\hat x^{\text{in}}, \hat{x}_1^\text{squ}, \hat{x}_2^\text{squ} , \hat{x}_3^\text{squ}, \hat p^{\text{in}}, \hat{p}_1^\text{squ},  \hat{p}_2^\text{squ}, \hat{p}_4^\text{squ})   \\
\hat{p}^{\text{aux}}_3 \hspace{-0.1cm} &=& \hspace{-0.1cm} \hat{p}^{\text{aux}}_3 (\hat x^{\text{in}}, \hat{x}_1^\text{squ}, \hat{x}_2^\text{squ} , \hat{x}_3^\text{squ}, \hat p^{\text{in}}, \hat{p}_1^\text{squ},  \hat{p}_2^\text{squ}, \hat{p}_4^\text{squ})   \\
\hat{p}^{\text{out}} \hspace{-0.1cm} &=& \hspace{-0.1cm} \hat{p}^{\text{out}}  (\hat x^{\text{in}}, \hat{x}_1^\text{squ}, \hat{x}_2^\text{squ} , \hat{x}_3^\text{squ}, \hat p^{\text{in}}, \hat{p}_1^\text{squ},  \hat{p}_2^\text{squ}, \hat{p}_4^\text{squ})  
\end{array} \right.%
\ea
where the functional expression depend on the applied transformation $U$.

Suppose now the quadrature $\hat p$ is measured on all the modes, except the last one, which represents the result of the computation, and whose measurement constitutes the readout. 
In the Heisenberg representation, the projective measurement of $\hat{p}^{\text{aux}}_i$ with $i = 1,2,3$ effectively results in replacing these operators by the corresponding measurement outcomes $p^{\text{aux}}_i$ in Eq.(\ref{eq:modi_sys2}), which are real numbers~\cite{Furusawa_Book}. Then, the linear system composed of the first $3$ lines in Eq.(\ref{eq:modi_sys2}) is solved for the anti-squeezed observables $\hat{x}_1^\text{squ},\hat{x}_2^\text{squ},\hat{x}_3^\text{squ}$. These are then replaced in the last line of Eqs.(\ref{eq:modi_sys1}) and (\ref{eq:modi_sys2}), i.e. in the expression of the output mode variables $\hat{x}^\text{out},\hat{p}^\text{out}$, yielding the result 
\ba
\label{eq:transf-single-mode}
\hat{x}^{\text{out}} = \sum_{i = 1}^{m} c_x^i \hat{p}^{\text{squ}}_i + a \hat{x}^{\text{in}} +  b  \hat{p}^{\text{in}}   \hspace{-0.05cm} + \hspace{-0.05cm}  \sum_{i = 1}^{m} l_{x}^i p^{\text{aux}}_i  \label{eq:trasf-simpl-effX} \\
\hat{p}^{\text{out}} = \sum_{i = 1}^{m} c_p^i \hat{p}^{\text{squ}}_i + c \hat{x}^{\text{in}} +  d \hat{p}^{\text{in}}  \hspace{-0.05cm} + \hspace{-0.05cm}  \sum_{i = 1}^{m} l_p^i p^{\text{aux}}_i.   \hspace{-0.1cm}  \label{eq:trasf-simpl-effP}
\ea
The coefficients $c_{x,p}^i, a,b,c, d$, and $l_{x,p}^i$ depend upon the specific transformation $U$ that is applied to the input modes according to Eq.(\ref{eqtrasf-y0}). The terms $ \sum_{i = 1}^{m} l_{x,p}^i p^{\text{aux}}_i $ are linear functions of the measurement outcomes that, although they may be corrected for, do not affect the symplectic structure of the input-output transformation~\cite{Nielsen2006,Gu2009}. The output mode of Eqs.(\ref{eq:trasf-simpl-effX}),(\ref{eq:trasf-simpl-effP}) encodes the result of the QC, and its measurement  (which can be simultaneous with the others) provides the result.
As already mentioned in the main text, this very general scheme encompasses both the traditional MBQC scheme based on cluster states, as well as the newly proposed method based on the post-processing.

\section{Determination of a lower bound on the minimal number of ancillary squeezed states required to cover the full symplectic group with the Direct MBQC method}
\label{se:minimal}

A lower bound on the minimal number $m_{\text{min}}$ of ancillary squeezed states to be employed to cover the full symplectic group with the direct MBQC method can be determined with group theory arguments. The most general symplectic tranformations to which the $n$-mode input state is subjected is described by $2 n^2 + n$ degrees of freedom~\cite{dutta1995real}.
The choice of the LO phases $\vec{\varphi}$ modeled by the matrix  $\Delta_{\mathrm{LO}}$ amounts to $ n + m_{\text{min}}$ degrees of freedom. 
For the purpose of MBQC, the only relevant matrices $O$ are those that mix the readout mode (i.e., the mode that encode the result) with the others. 
Other rotation matrices would only affect the non-symplectic part of Eq.(\ref{eq:trasf-simpl-eff}), i.e. the displacements  ${\eta}_{x}^i $ and ${\eta}_{p}^i$. Each output mode is therefore mixed either with one of the ancillary measured modes, yielding $n \cdot m_{\text{min}}$ possible elementary rotations, or with another output mode, yielding the rotation group in dimension $n$, which is parameterized by $n (n -1)/2$ degrees of freedom. 

The relevant condition   is hence
\be
2 n^2 + n = (n + m_{\text{min}}) + n \cdot m_{\text{min}} + n (n -1)/2  
\ee
which yields as a solution
\be
m_{\text{min}} = \fr{3}{2} n,
\ee
i.e. a linear scaling of the minimal number ancillary modes with the number of modes of the input state to be processed.

\section{Details on the optimization procedure}
\label{se:how-optimization-is-performed}

We discuss here the details of the numerical  optimization procedure yielding the solution in Eq.(\ref{eq:teleport_result2}) for the Fourier transform of the input state (analogous considerations hold for the optimization of the $C_z$ gate).

The starting point is Eq.(\ref{eq:trtot_m}), which establishes the available transformations that can be performed on the input state and auxiliary modes. In the case considered of the Fourier transform we have a single-mode state that has to be transformed, and 3 auxiliary modes, for a total of 4 optical modes. 
The matrix $U_T$ is given in Fig.\ref{fig:UT}, modulo a re-labeling of the modes (which results in a shuffle of the matrix elements), as we chose to take as input state the state carried by the fourth mode of the mode basis in Fig.\ref{fig:UT} (left). This (unessential) choice is dictated by the fact that in the experiment we mainly refer to, Ref.~\cite{Roslund_13b}, the fourth mode carries the less squeezed state, and therefore the three squeezed states carried by the remaining optical modes provide a better auxiliary resource. 
The matrix $\Delta_{\mathrm{LO}} (\vec{\varphi}) = \text{diag}  \lgr e^{i \varphi_1}, e^{i \varphi_2}, e^{i \varphi_3}, e^{i \varphi_4} \rgr $ brings 4 degrees of freedom that can be used for our optimization.  
The orthogonal transformation $O (\vec{\theta})$ can be parameterized in terms of elementary rotations, e.g. in terms of the Tait-Bryan angles. 
As mentioned in Sec.\ref{se:minimal}, the relevant orthogonal transformations are only those that mix the output mode with the auxiliary modes, and not those which only mix auxiliary modes among each other. In this case, therefore, only three elementary transformations are needed, correspondingly parameterized by three angles $\theta_1, \theta_2, \theta_3$.
Each choice of these $7$ available angles $(\vec{\theta}, \vec{\varphi})$ corresponds to a total unitary transformation Eq.(\ref{eq:trtot_m}). For each such trial unitary we compute the output quadratures as specified by Eqs.(\ref{eq:modi_sys1}, \ref{eq:modi_sys2}), and the output modes in Eq.(\ref{eq:transf-single-mode}), following the mathematical procedure explained in Appendix \ref{eq:app-mbqc}.
This determines the coefficients $a(\vec{\theta}, \vec{\varphi}),b(\vec{\theta}, \vec{\varphi}),c(\vec{\theta}, \vec{\varphi}),d(\vec{\theta}, \vec{\varphi})$ associated to the trial angles. Our program then evaluates the Frobenius matrix-distance between the corresponding symplectic matrix to the target one. In reference to Eq.(\ref{eq:fitness-last}), in the case of  the Fourier transform we have
\be
\label{eq:fitness-last-app}
f_{1} (\vec{\theta}, \vec{\varphi}) = \| \left( \begin{array}{cccccccc}
a(\vec{\theta}, \vec{\varphi}) & b(\vec{\theta}, \vec{\varphi})  \\
c(\vec{\theta}, \vec{\varphi}) & d(\vec{\theta}, \vec{\varphi}) \\
\end{array} \right) - \left( \begin{array}{cccccccc}
0 & -1  \\
1 & 0 \\
\end{array} \right) \|.
\ee
In order to discover, among all the possible choices of the angles $(\vec{\theta}, \vec{\varphi})$, the one that minimizes Eq.(\ref{eq:fitness-last-app}), we use an evolutionary strategy as the one that was developed in Ref.\cite{roslund2009accelerated} by one of the authors of this work. Loosely speaking, these algorithms mimic Darwinian evolution in order to find the solution that minimizes the ``fitness" function $f_{1}$. 
An iteration of the algorithm is called a generation. At each generation the algorithm starts from a point in the parameters space. At the first generation the starting point is chosen at random. Several points, called ``mutants", are then randomly generated around the starting one. The probability distribution of mutations is Gaussian and isotropic at the first iteration. The fitness function is evaluated for each of the mutants, which are then ranked according to the respective fitness. Half of the mutants (the ones with highest fitness) are linearly combined to generate a new point, which will be the starting point of the next iteration. Statistical analysis is carried out on the best mutants and the result is used to adjust the Gaussian probability distribution of mutants at the next iteration in order to speed up the convergence.
The sequence of generations is continued until after that 5000 loops are accomplished.
Once a solution is found, we compute the associated excess noise terms, yielding the noise values reported in Tab.~\ref{table-QC}. It turns out that, for the considered problem, the excess noise is conspicuously reduced for the solution found optimizing $f_1$ (compared to the standard procedure consisting in generating and measuring a cluster state from the same auxiliary state), even though $f_1$ does not depend on the excess noise, whose reduction was thus not directly addressed by the optimization. The problem of explicitly reducing the excess noise could be tackled either modifying the fitness function, for example subtracting the excess noise to $f_1$, or by performing a multi-objective optimization strategy. In other words, the nice noise reduction we found was collateral, but it could be systematically addressed by means of refined optimization routines when dealing with operations for which this would be needed.


%
\section{Experimental matrix $U_T$ used in the $C_Z$ gate optimization}
\label{se:exp-values}

As mentioned in the main text, in order to study the implementation of a $C_Z$ gate we refer to a six-mode run of the experiment reported in Ref~\cite{Roslund_13b}. There, six independent squeezed states are produced and accessed in an optical cavity by Spontaneous Parametric Down Conversion with a femto-second pump. The change of basis implemented by detecting the light beam in a 6-mode pixel basis is described by the matrix
\small{ \be
U_T \hspace{-0.05cm} = \hspace{-0.05cm}
\left(
\begin{array}{cccccc}
 -0.45 & -0.619 & 0.536 & 0.334 & -0.124 & -0.00859 \\
 -0.363 & -0.326 & -0.161 & -0.635 & 0.521 & 0.246 \\
 -0.334 & -0.133 & -0.383 & -0.248 & -0.402 & -0.708 \\
 -0.326 & 0.0013 & -0.466 & 0.143 & -0.498 & 0.639 \\
 -0.365 & 0.155 & -0.382 & 0.607 & 0.547 & -0.174 \\
 -0.561 & 0.685 & 0.421 & -0.187 & -0.0645 & 0.0107 \\
\end{array}
\right). \nn
\ee}
\vspace{0.5cm}

As two input states on which operate the gate we can consider the first two squeezed modes of the cavity, i.e. two independent squeezed states. The experimental implementation in Ref.~\cite{Roslund_13b} displays alternating squeezing quadratures between $x$ and $p$. Therefore, the transformation $U_T$ given above is multiplied from the right by $\Delta_{\mathrm{OPO}} = \text{diag} (1, 1, i, 1 , i, 1)$ in order encompass in the modelization of the accessible operations this more general case of ancillary squeezed states which are not all squeezed onto the same quadrature. Indeed, the effect of $\Delta_{\mathrm{OPO}}$ is to align the squeezing quadrature to $\hat p$ for the four last squeezed modes, to serve as the ancillary resource, thereby matching the input state of Eq.(\ref{eq:trtot_m}). This corresponds to having as an input state of the $C_Z$ gate a couple of independently squeezed states, one on $x$ and the other on $p$.

\bibliographystyle{apsrev}
\bibliography{multimode-BIB}

\end{document}